\documentclass[twocolumn,showpacs,aps,prd,superscriptaddress]{revtex4}

\usepackage{graphicx}
\usepackage{dcolumn}
\usepackage{amsmath}
\usepackage{epsfig}

\usepackage{times}
\usepackage{here}
\usepackage{lscape}
\usepackage{rotating}
\usepackage{wrapfig}
\usepackage{psfrag}
\usepackage{color}
\usepackage{rotate}
\usepackage{amssymb}
\usepackage{multirow}

\input babarsym

\newcommand{\BABARPubYear}    {07}
\newcommand{\BABARPubNumber}  {068}

\newcommand{\SLACPubNumber} {13047}

\def\figurebox#1#2#3{%
    \def\arg{#3}%
    \ifx\arg\empty
    {\hfill\vbox{\hsize#2\hrule\hbox to #2{\vrule\hfill\vbox to
     #1{\hsize#2\vfill}\vrule}\hrule}\hfill}%
    \else
    {\hfill\epsfbox{#3}\hfill}%
    \fi}

\long\def\inst#1{\par\nobreak\kern 4pt\nobreak
    {\it #1}\par\vskip 10pt plus 3pt minus 3pt}

\begin{document}

\begin{flushleft}

\babar-PUB-\BABARPubYear/\BABARPubNumber\\
SLAC-PUB-\SLACPubNumber\\
\end{flushleft}

\title{
\Large \bf \boldmath
Measurement of $\Dz$-$\Dzb$ Mixing using the Ratio of Lifetimes for the Decays
$\Dz \to K^-\pi^+$, $K^-K^+$, and $\pi^-\pi^+$
}

%
\author{B.~Aubert}
\author{M.~Bona}
\author{Y.~Karyotakis}
\author{J.~P.~Lees}
\author{V.~Poireau}
\author{X.~Prudent}
\author{V.~Tisserand}
\author{A.~Zghiche}
\affiliation{Laboratoire de Physique des Particules, IN2P3/CNRS et Universit\'e de Savoie, F-74941 Annecy-Le-Vieux, France }
\author{J.~Garra~Tico}
\author{E.~Grauges}
\affiliation{Universitat de Barcelona, Facultat de Fisica, Departament ECM, E-08028 Barcelona, Spain }
% ...
\collaboration{The \babar\ Collaboration}
\noaffiliation

\date{\today}

\begin{abstract}

We present a measurement of $\Dz$-$\Dzb$ mixing parameters using the ratios of
lifetimes extracted from a sample of \Dz mesons produced through the process
$\Dstp\to\Dz\pip$, that decay to \Kmpip,
\KmKp, or \pipi. The Cabibbo-suppressed modes \KmKp and \pipi are
compared to the Cabibbo-favored mode \Kmpip to obtain a measurement
of \yCP, which in the limit of \CP conservation corresponds to
the mixing parameter $y$. 
The analysis is based on a data sample of 384~\invfb
collected by the \babar\ detector at the PEP-II asymmetric-energy
\epem collider. We obtain
$  \yCP     =  [1.24\pm 0.39\stat \pm 0.13\syst]\%$, which is evidence
of \Dz-\Dzb mixing at the $3\sigma$ level, and $\deltaY  =  [-0.26\pm
0.36\stat \pm 0.08\syst]\%$, where $\deltaY$ constrains possible \CP{} violation.
Combining this result with a previous \babar{} measurement of \yCP
obtained from a separate sample
of $\Dz\to\KmKp$ events, we obtain
$  \yCP     =  [1.03\pm 0.33\stat \pm 0.19\syst]\%$.

\end{abstract}
\pacs{13.25.Ft, 12.15.Ff, 11.30.Er}
\maketitle

\label{sec:intro}
Several recent studies have shown evidence for mixing in
the \Dz-\Dzb{} system at the 1\% 
level~\cite{Aubert:2007wf,Staric:2007dt,Abe:2007rd}. 
This is consistent with Standard Model (SM)
expectations~\cite{foundations} and provides strong constraints for new physics
models~\cite{surveys}.
One consequence of \Dz-\Dzb{} mixing is that the \Dz\ decay time
distribution can be different for decays to different \CP
eigenstates~\cite{Liu:1994ea}. An observation of 
\CP violation in \Dz-\Dzb{} mixing with the present experimental sensitivity would provide evidence for physics
beyond the SM~\cite{Blaylock:1995ay}. We present a
measurement of this lifetime difference and results of a search for
evidence of \CP violation in \Dz-\Dzb mixing.

The two neutral $D$ mass eigenstates $| D_1 \rangle$ and $| D_2
\rangle$  can be represented as 
\begin{equation}
\begin{array}{rcl}
| D_1 \rangle &=& p | D^0 \rangle + q | \Dzb \rangle \\
| D_2 \rangle &=& p | D^0 \rangle - q | \Dzb \rangle \;,
\end{array}
\label{eq:qpdef}
\end{equation}
where $\left|p\right|^2 + \left|q\right|^2 = 1$.  We characterize the
rate of $D^0$-$\Dzb$ mixing with the parameters $x \equiv \Delta m/\Gamma$ and $y \equiv \Delta\Gamma/2\Gamma$, where
$\Delta m = m_1 - m_2$ and $\Delta \Gamma = \Gamma_1 - \Gamma_2$ 
are the differences between the mass and width eigenvalues of the 
states in Eq.~(\ref{eq:qpdef}), respectively, and
$\Gamma = (\Gamma_1+\Gamma_2)/2$ is the average width.  If either $x$
or $y$ is non-zero, mixing will occur.

The effects of \CP violation in $\Dz$-$\Dzb$ mixing
can be parameterized in terms of the quantities
\begin{equation}
r_m \equiv \left| \frac{q}{p} \right|
\mbox{\hskip 0.25in}{\rm and}\mbox{\hskip 0.25in}
\varphi_f \equiv \arg\left( \frac{q}{p}\frac{\overline{A}_f}{A_f} \right) \;,
\end{equation}
where $A_f \equiv \langle f | {\mathcal H}_D | D^0 \rangle$
($\overline{A}_f \equiv \langle f | {\mathcal H}_D | \Dzb
\rangle$) is the amplitude for $D^0$ ($\Dzb$) to decay into a
final state $f$, and ${\mathcal H}_D$ is the Hamiltonian for the decay. A value of $r_m \neq 1$ would indicate \CP
violation in mixing. A non-zero value of $\varphi_f$ would
indicate \CP violation in the interference between mixing and decay.
Within the SM, \CP violation in decay is expected to be small in the $D^0$-$\Dzb$
system~\cite{Bergmann:2000id} and is considered elsewhere~\cite{DirectCPVResults}.

$D^0$-$\Dzb$ mixing will alter the decay time distribution of
$D^0$ and $\Dzb$ mesons that decay into final states of specific
\CP. To a good approximation, these decay time distributions can
be treated as exponential with effective
lifetimes \tauhhp and \tauhhm, given by~\cite{Bergmann:2000id}
\begin{equation}
\begin{array}{rcl}
\tauhhp   &=&
\tauKpi\left[ 1 + r_m \left( y\cos\varphi_f -
                                 x\sin\varphi_f \right) \right]^{-1} \\
\tauhhm &=&
\tauKpi\left[ 1 + r_m^{-1} \left( y\cos\varphi_f +
                                      x\sin\varphi_f \right) \right]^{-1} \;,
\end{array}
\end{equation}
where $\tauKpi$ is the lifetime for the Cabibbo-favored decays
$D^0 \to K^-\pi^+$ and $\Dzb \to K^+\pi^-$, and $\tauhhp$ ($\tauhhm$)
is the lifetime for the Cabibbo-suppressed decays
of the $D^0$ ($\Dzb$) into \CP-even final states (such as $K^-K^+$
and $\pi^-\pi^+$).
These effective lifetimes can be combined into the
 quantities $\yCP$ and $\Delta Y$:
\begin{equation}
\begin{array}{rcl}

\displaystyle\yCP = \frac{\tauKpi}{\langle\tauhh\rangle} - 1\\[0.5cm]

\displaystyle\Delta Y = \frac{\tauKpi}{\langle\tauhh\rangle} A_\tau \;,
\label{eq:yCPcalc}
\end{array}
\end{equation}
where $\langle \tauhh \rangle = (\tauhhp+\tauhhm)/2$ and
$A_\tau = (\tauhhp-\tauhhm)/(\tauhhp+\tauhhm)$. Both $\yCP$ and $\deltaY$
are zero if there is no $D^0$-$\Dzb$ mixing. 
In the limit of \CP conservation, $\yCP = y$ and $\deltaY = 0$,
with the convention that $\cos\varphi_f > 0$.

\label{sec:strategy}

We measure the \Dz lifetime in
the three different \Dz decay modes, \Kmpip, \KmKp, and \pipi. We
use \Dz mesons coming from $\Dstp\to\Dz\pip$ 
decays~\cite{footnote:CC}; the requirement of a \Dstp{} parent
strongly suppresses the backgrounds. We use the charge of the
$D^{*\pm}$ to split the \KmKp and \pipi samples into those originating
from \Dz and from \Dzb
mesons for measuring the \CP-violating parameters.
To avoid potential bias, we finalize our data selection criteria, the procedures for
fitting and for extracting the statistical limits, and determine the
systematic errors, prior to examining the mixing results.

Most systematic errors related to signal events are expected to cancel in
the lifetime ratios. Background events can contain effects that
differ in each decay mode, making them difficult to characterize.  Therefore,
the event selection is chosen to produce very pure samples.
The decay time distribution of signal candidates is fit to
an exponential convolved with a resolution function that uses
event-by-event decay time errors. The decay-time resolution parameters
are allowed to vary in the fit.  Residual background components are
modeled using Monte Carlo (MC) simulated events and control samples 
obtained from the data.

\label{sec:event}

We use $384\invfb$ of \epem\ colliding-beam data recorded near
$\sqrt{s} = 10.6\gev$ with the \babar\ detector~\cite{Aubert:2001tu}
at the PEP-II asymmetric-energy storage rings.
We begin by reconstructing candidate \Dz decays into the final states
$\Kmpip$, \pipi, and \KmKp. We require tracks to satisfy
particle identification criteria based upon \dedx ionization energy loss 
and Cherenkov angle measurements.  
We fit pairs of tracks with the appropriate
mass hypotheses to a common vertex.
We require the invariant mass of a candidate track pair to be within 
the range 1.78--1.94~\gevcc.
To further reduce backgrounds, we require the
helicity angle $\theta_H$, defined as the angle between the positively
charged track in the \Dz rest frame and the \Dz direction in the lab frame, to
satisfy $|\cos\theta_H|<0.7$. This is particularly helpful for 
rejecting combinatorial background, especially in the \pipi mode.

We reconstruct \Dstarp candidates by combining a \Dz candidate with a 
slow pion track (denoted $\pisoft^+$), requiring them to originate 
from a common vertex constrained to the \epem interaction region.  
We require the $\pisoft^+$ momentum
to be greater than 0.1 \gevc in the laboratory frame and less than
0.45 \gevc in the \epem center-of-mass (CM) frame.
We perform a vertex-constrained combined fit to the \Dz production 
and decay vertices, requiring 
the \chisq-based probability, $\Pchisq$, to be at least $0.1\%$.
The decay time \t\ and its estimated uncertainty~\terr\
for each \Dz candidate are determined by this fit.
We reject slow electrons that fake $\pisoft^+$ candidates
using \dedx measurements in the tracking volume and further veto  
any $\pisoft^+$ candidate that may have originated from a reconstructed
gamma conversion or \piz Dalitz decay.

To reduce combinatorial backgrounds from
\Dz's produced via $B$-meson decay we require each \Dz to have a momentum in
the CM frame greater than $2.5\gevc$. We also require $-2 < \t<4\ps$ and
$\terr<0.5\ps$.  The most probable value of \terr for signal events is
$0.16\ps$.  For cases where multiple $D^{*\pm}$ candidates in an event
share one or more tracks, we retain only the candidate with the highest
\Pchisq.

\begin{figure}[!ht]
\hbox to \hsize{
 \includegraphics[width=0.5\linewidth]{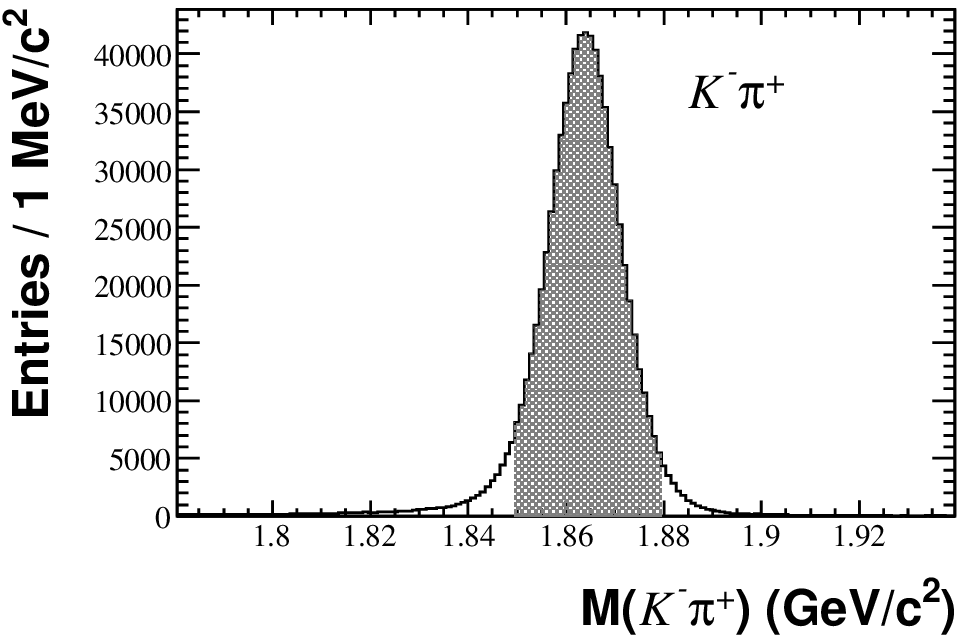}\hfill
 \includegraphics[width=0.5\linewidth]{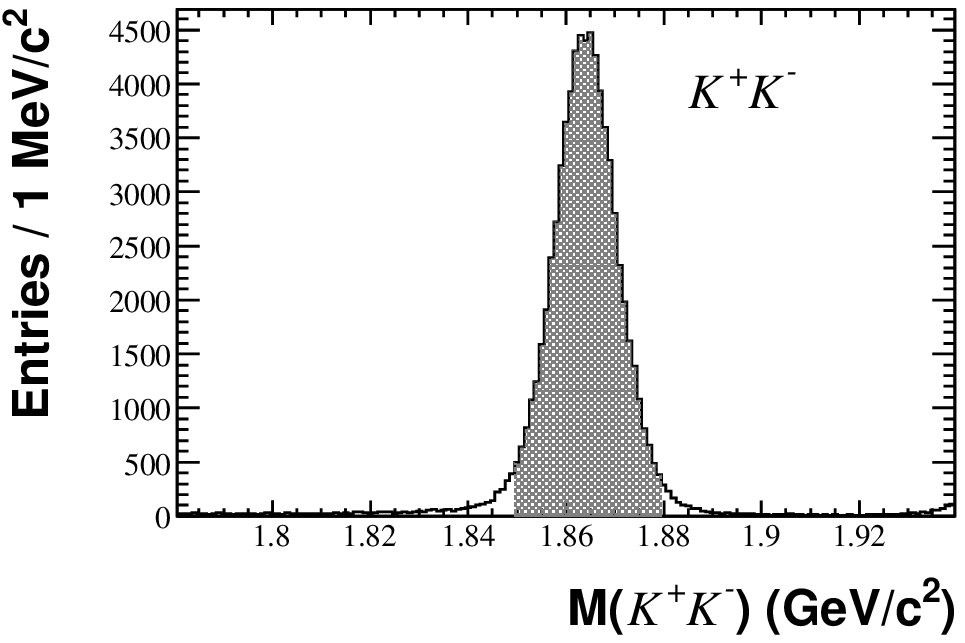}
}
\hbox to \hsize{
 \includegraphics[width=0.5\linewidth]{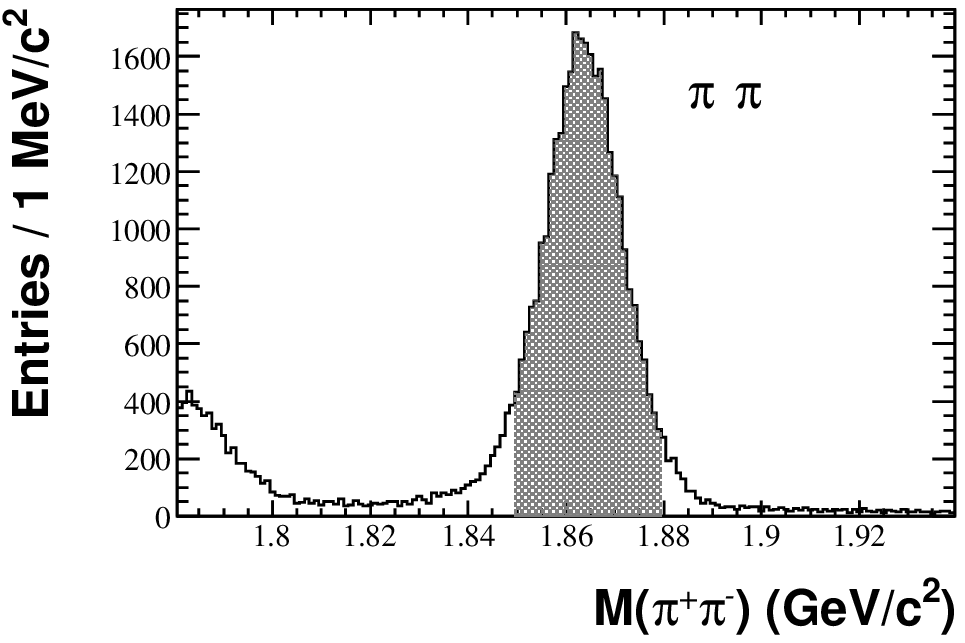}\hfill
\raise1.5cm\hbox{\begin{tabular}{lcr}\hline\hline
Sample &  Size & Purity (\%) \\
\hline
$K^-\pi^+$         & 730,880 & 99.9  \\
$K^-K^+$           &  69,696 & 99.6  \\
$\pi^-\pi^+$       &  30,679 & 98.0  \\
\hline\hline
\end{tabular}}
}
\caption{\label{fig:MassPlot}The reconstructed $D^0$ mass distributions
for the three $D^0$ samples, within $\pm$0.8~\mevcc of the peak of $\delta
m$. The shaded region indicates the events used
in the lifetime fit. (The structures appearing
above 1.92 \gevcc in the \KmKp decay mode, and below 1.81 \gevcc in the
\pipi decay mode, are mainly due to candidates
with misidentified kaons or pions.) Also shown are the yield and purity of the three $D^0$ samples
as calculated inside the $\pm 15\mevcc$ mass window.}
\end{figure}

The distribution of the difference in the reconstructed $D^{*+}$ and
$D^0$ masses ($\delta m$) peaks near $145.4$~\mevcc.  Backgrounds are
suppressed by discarding $D^{*+}$ candidates with a value of $\delta
m$ deviating more than 0.8~\mevcc from the peak. Invariant mass
distributions for the selected $D^0$ candidates are shown in
Fig.~\ref{fig:MassPlot}. For the lifetime fit, we only use events
within $15\mevcc$ of the \Dz signal peak (shaded regions
in Fig 1); the event yields and purity
within this signal region are also given.

\label{sec:fit}


The \Dz lifetime is determined from an unbinned maximum likelihood fit
to the reconstructed decay time and its estimated error for events in the
signal region. The fit is performed simultaneously to all five decay samples
(\Dz\to\Km\Kp; \Dzb\to\Kp\Km; \Dz\to\pim\pip; \Dzb\to\pip\pim;
$\Dz\to\Km\pip$ and $\Dzb\to\Kp\pim$ combined).
The \Dz candidates in the signal region can be divided into three
components: \Dz signal events, combinatorial background, and
mis-reconstructed charm events.  Each component is described by its own
probability density function (PDF) which also depends upon the \Dz or
\Dzb decay mode.

The measured decay-time distribution of signal events is
described by an exponential convolved with a resolution function.  The
resolution function is the sum of three Gaussian functions with widths
proportional to \terr. The three Gaussian functions share a common mean which is
allowed to be offset from zero in order to take detector misalignment
effects into account. The effect of the offset is studied as part of the
crosschecks and taken into account as a systematic uncertainty.  The resolution function
parameters are all permitted to vary in the fit.
Up to an overall scale factor in the width, the resolution function is 
observed to have the same shape for all modes, including the offset. To 
account for the small (1.5\%) differences in width, we introduce two parameters $S_{\KmKp}$ and 
$S_{\pipi}$ to scale the overall width of the \KmKp and \pipi resolution functions 
relative to the width of the $\Km\pip$ resolution function. All other resolution 
function parameters are shared among the different modes and are 
determined by a simultaneous fit to all modes together.

About $0.4\%$
of the \Dz signal in the \KmKp and \pipi modes consists of a correctly
reconstructed \Dz combined with an unrelated $\pisoft$; this is
estimated from MC and verified in data.  These
candidates have the same
resolution and lifetime behavior as those from correctly reconstructed \Dstp{}
decays, but about half of them will be tagged as the wrong
flavor. We therefore include a $0.2\%$ component in the signal PDF that uses
the lifetime of the opposite flavor state.

The decay-time distribution of the combinatorial background is
described by a sum of a Gaussian and a modified Gaussian with a power-law tail
to account for a small number of events with large reconstructed lifetimes.
The means of these functions are allowed to float in the fit. Each of
the three decay modes has its own
shape for the combinatorial background. These shapes are determined from fits
to the events in the sideband region defined by $1.89<\m_{hh}<1.92\gevcc$ and
$0.151<\delta m<0.159\gevcc$. We determine the amount of combinatorial background using MC samples scaled to the same luminosity as the
data, modeling all known, relevant physics processes. The fraction of combinatorial background in the \Kmpip mode is
estimated to be $(0.032\pm0.003)\%$, in the \KmKp mode $(0.16\pm0.02)\%$,
and in the \pipi mode $(1.8\pm0.2)\%$. The uncertainties are
determined by comparing data and MC events in the
$(\m_{hh},\delta m)$ sideband where the combinatorial background is
dominant.

Mis-reconstructed charm background events have one or more of the charm decay
products either not reconstructed or reconstructed with the wrong
particle hypothesis. Most are \Dz mesons from a $\Dstp\to\Dz\pisoft$
decay with a correctly reconstructed \pisoft. For the \Kmpip mode, most of
the charm background is semileptonic decays $\Dz\to\Km\ell^+\nu$ with
the charged lepton misidentified as a pion. The semileptonic decays
also contribute to the \KmKp final state, but the dominant contribution is
from $\Dz\to\Kmpip\piz$ in which the \piz is not reconstructed
and the \pip is misidentified as a kaon. There is also a small
contribution from $\Dp\to\Kmpip\pip$ decays. In the \pipi mode, the
charm background is almost exclusively due to mis-reconstructed $\Dz\to\Km\pip$ decays in which the kaon has been
misidentified as a pion. The decay-time distributions of the charm backgrounds are
described by an exponential convolved with a Gaussian. The parameters are
fixed to values 
obtained in a fit to MC events. The fraction of charm background
events in the signal region is 
estimated from MC simulation and crosschecked by comparing data and MC
events in a
$(\m_{hh},\delta m)$ sideband region defined by $1.78<\m_{hh}<1.80\gevcc$ and
$0.14<\delta m<0.16\gevcc$, where the charm background is the dominant 
contribution. We estimate the charm background to be
$(0.009\pm0.002)\%$ of events in the signal region for  \Kmpip,
$(0.2\pm0.1)\%$ for \KmKp, and $(0.15\pm0.15)\%$ for \pipi.

\begin{figure}[phtb]
  \centerline{
    \includegraphics[width=0.45\linewidth, clip=]{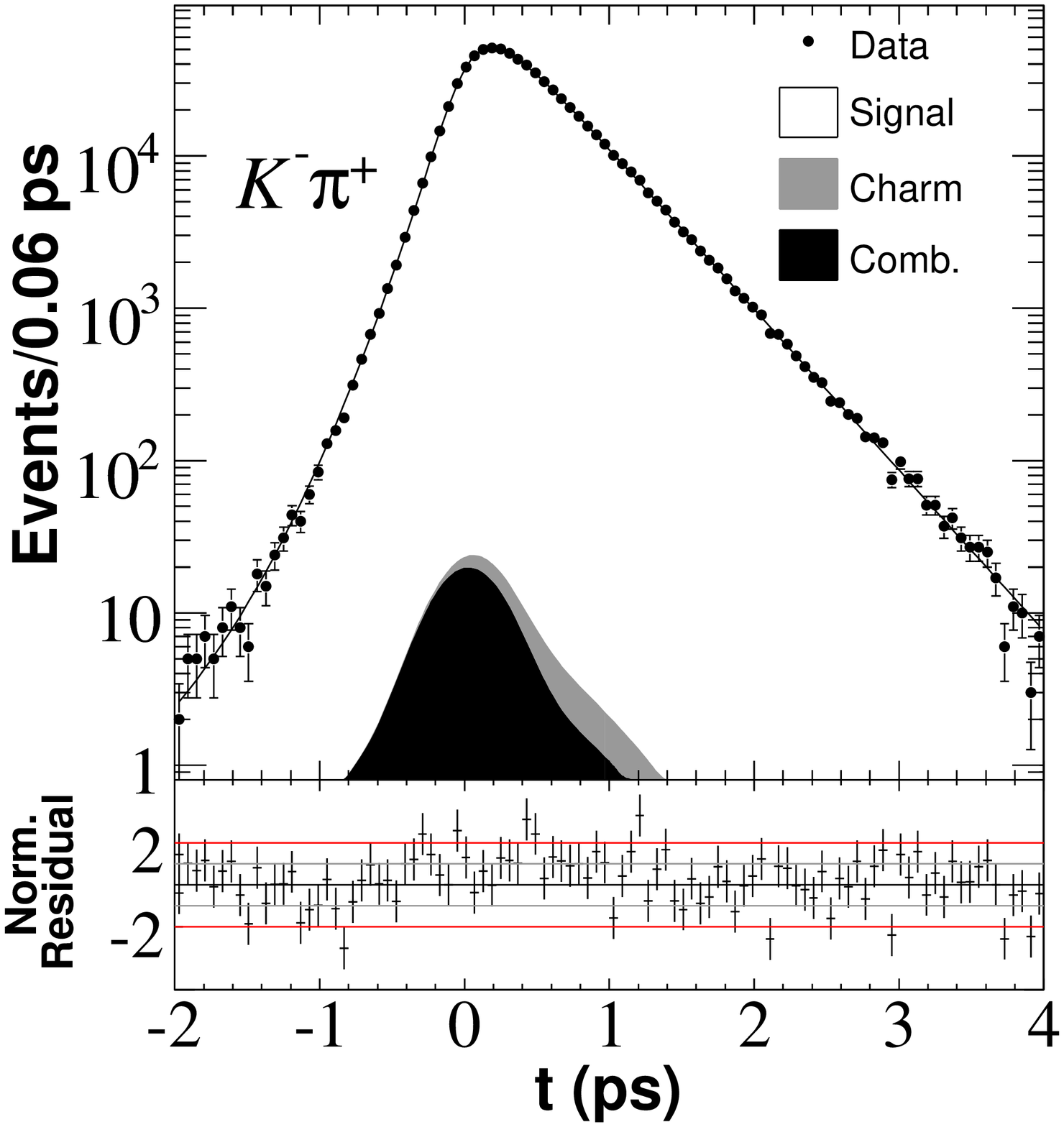}
    \raise 0.01cm \hbox{\includegraphics[width=0.45\linewidth, clip=]{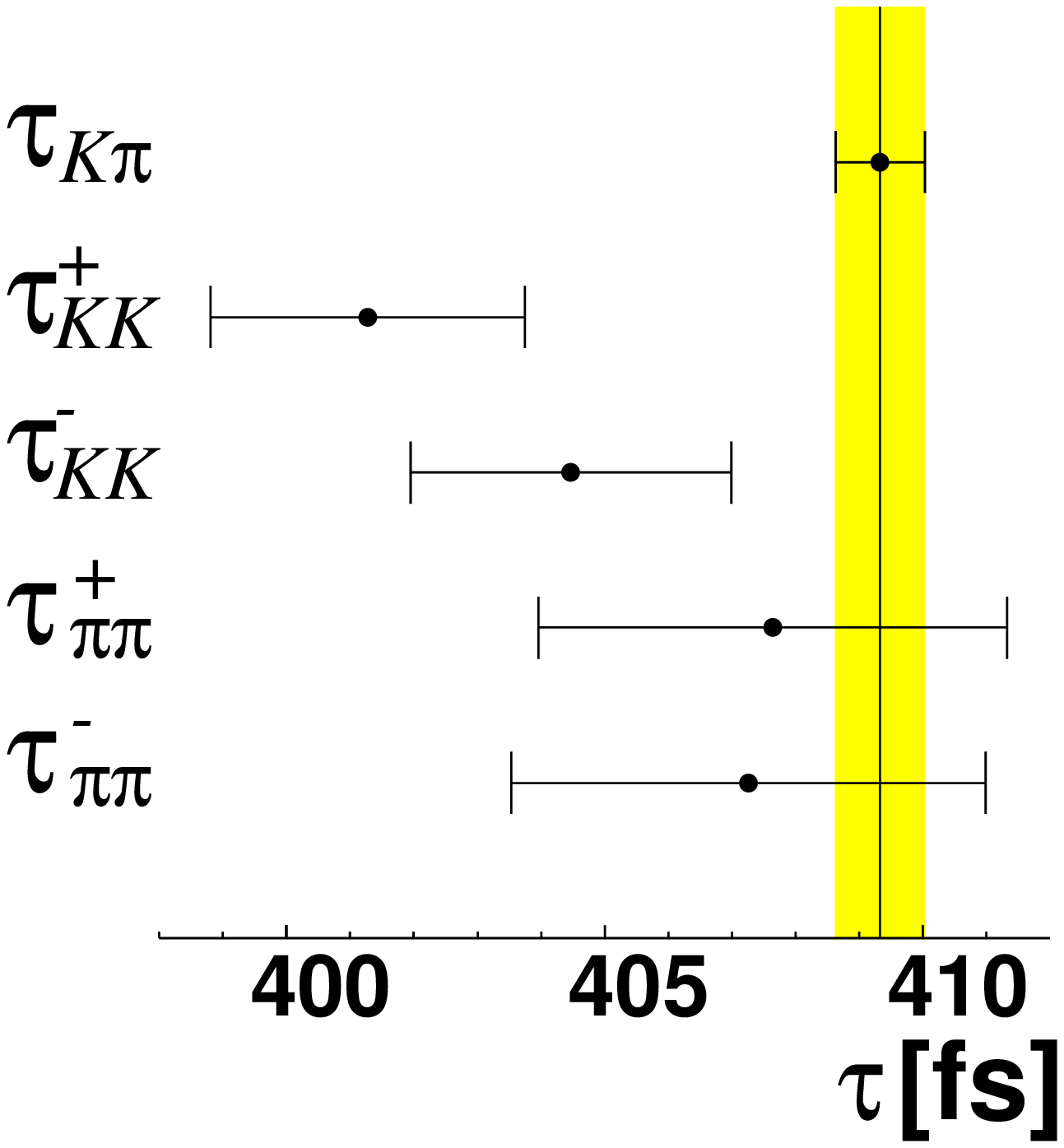}}
  }
  \centerline{%
    \includegraphics[width=0.45\linewidth, clip=]{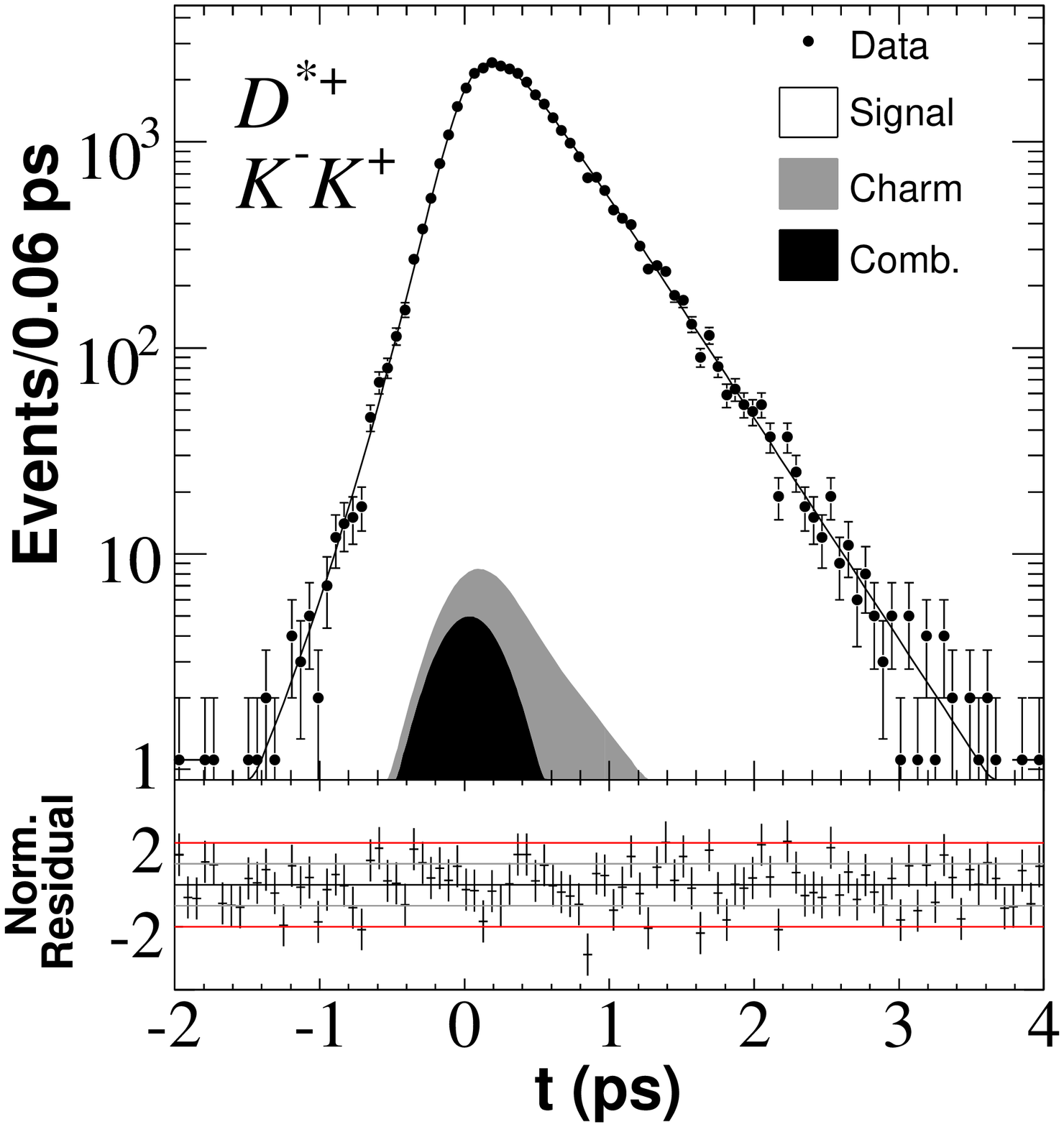}
    \includegraphics[width=0.45\linewidth, clip=]{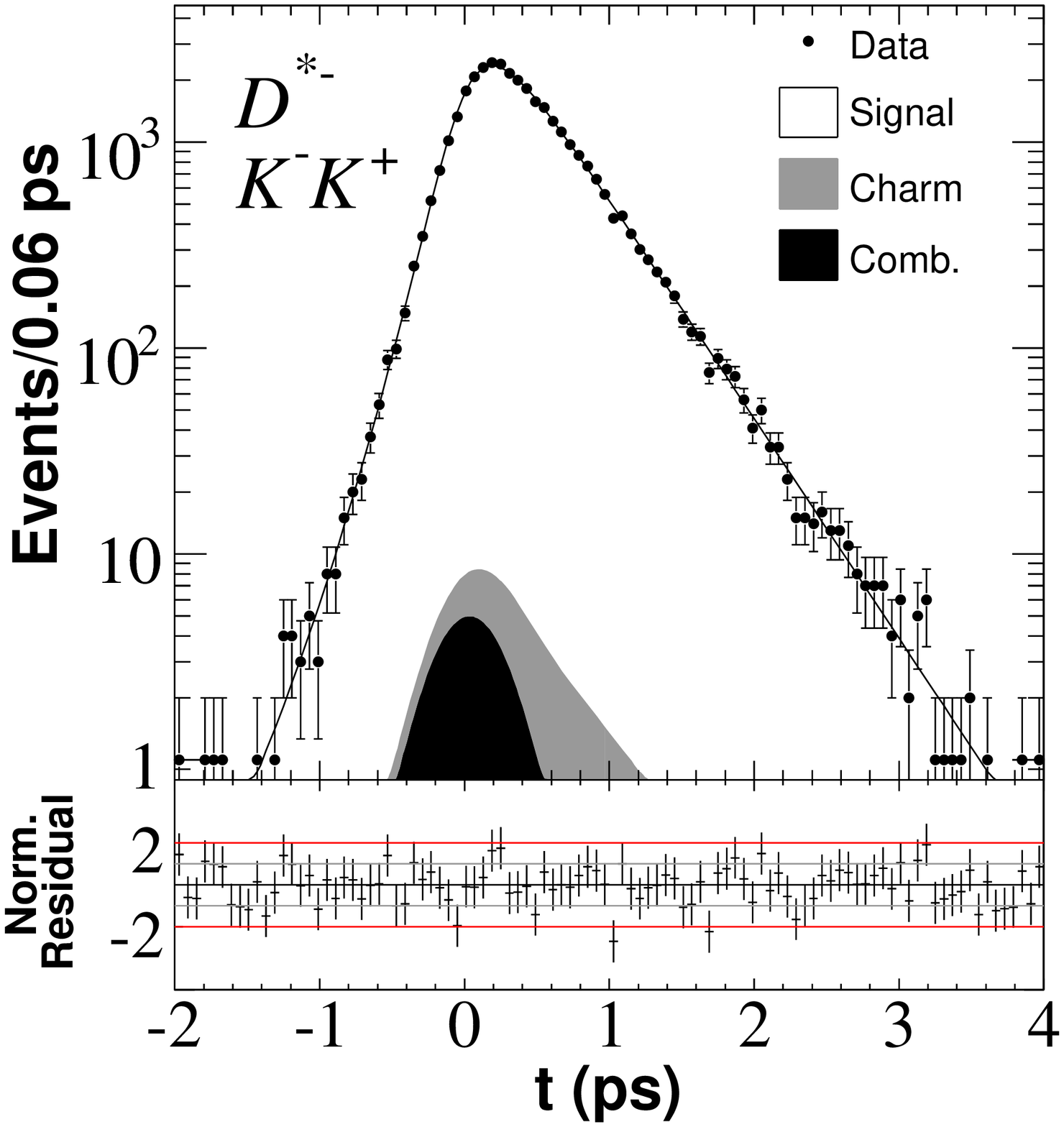}
  }
  \centerline{%
    \includegraphics[width=0.45\linewidth, clip=]{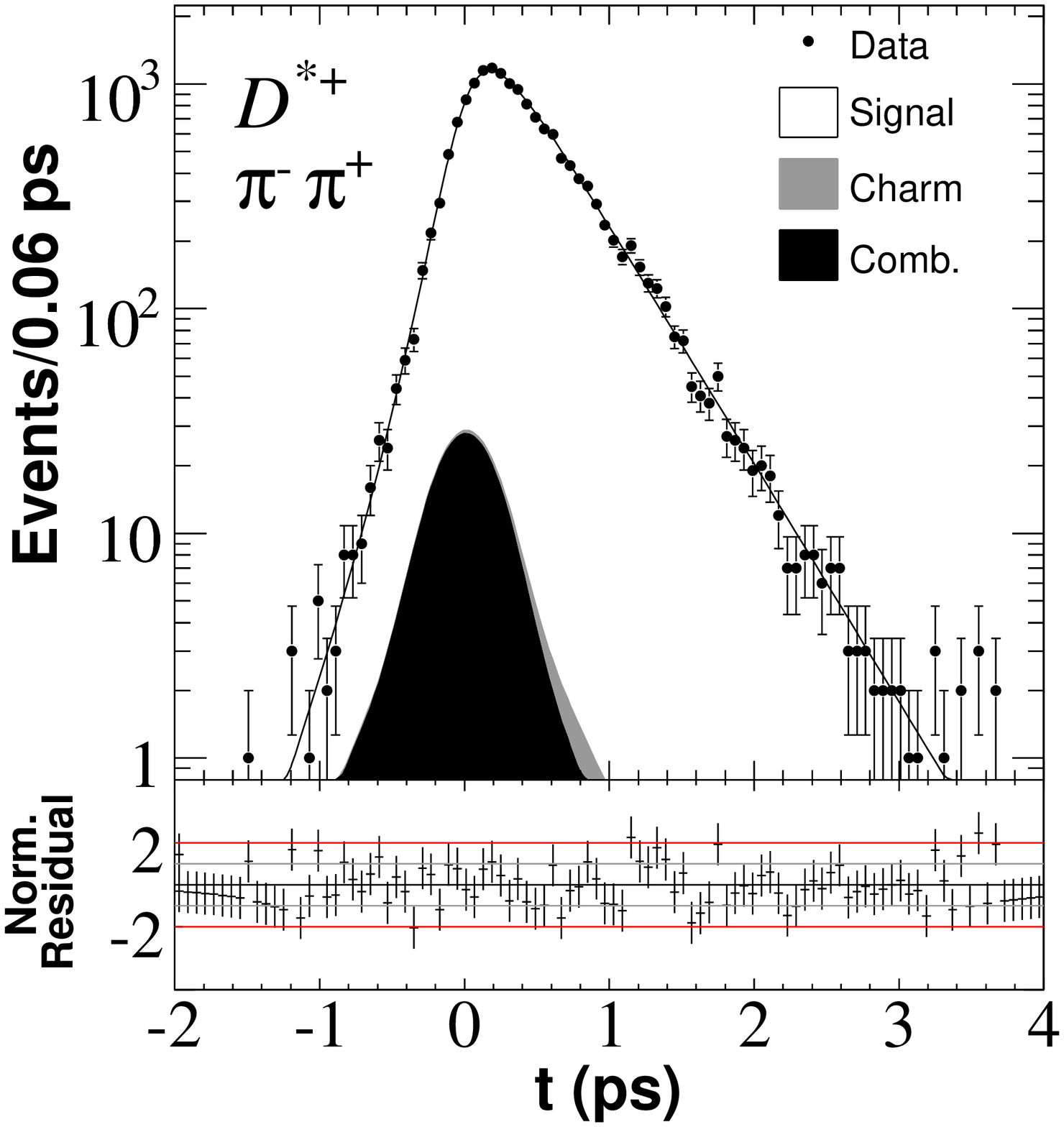}
    \includegraphics[width=0.45\linewidth, clip=]{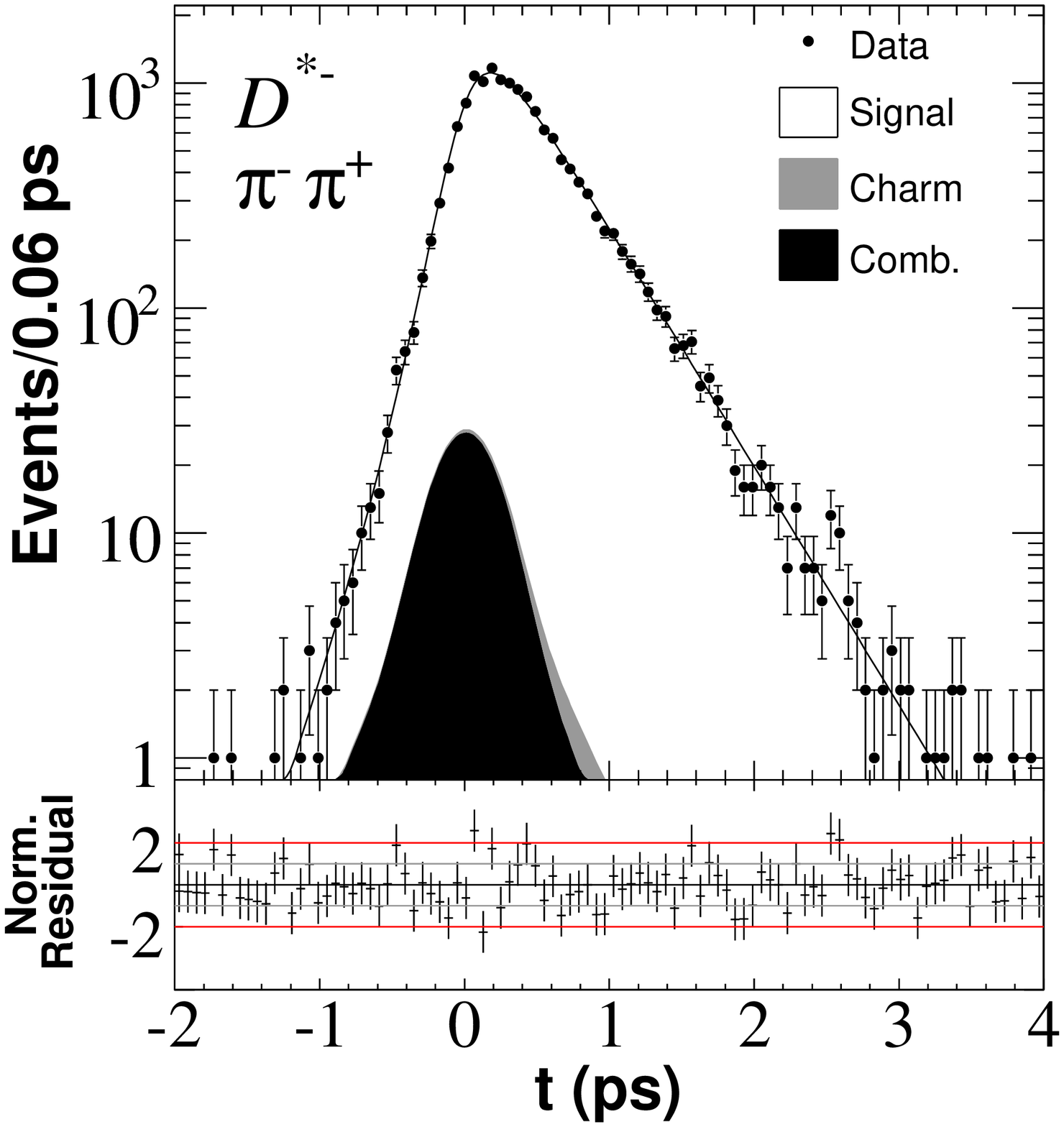}
  }
  \caption{Decay time distribution in the data samples with the combined fit overlaid.
           The top left plot is the  tagged \Kmpip sample, the middle plots are the \Dstp{} (left) and \Dstm{} (right) tagged \KmKp samples,
           and the bottom plots are the  tagged \pipi samples. The shaded and
           black distributions represent the charm and combinatorial background
   in the fit, respectively. The normalized residuals for each fit are shown
    as a separate histogram for each sample. The top right plot shows
    a summary of the measured lifetimes.}
  \label{fig:DecayTime}
\end{figure}

The results of the lifetime fits are shown in
Fig.~\ref{fig:DecayTime}. The fitted
\Dz lifetime $\tau_{K\pi}$  is found to be $409.33\pm0.70$ \stat \fs, consistent with the world-average
lifetime~\cite{Yao:2006px}.
From the fit results we calculate \yCP and \deltaY
for the \KmKp mode, the \pipi mode, and the two
modes combined,  taking into account any correlations
between the fitted lifetimes. The dominant correlation of 11\% arises
primarily because the decay-time resolution offset is shared between the decay
modes. The \yCP and \deltaY results are listed in
Table~\ref{tab:Results}.  The combined result is obtained by fitting the
data with common lifetimes for the \KmKp and \pipi modes, and assuming the same value of $\varphi_f$ for the $K^-K^+$ and
$\pi^-\pi^+$ decay modes.

\begin{table}[thb]
    \caption{The mixing parameters extracted from the fit to data,
    where the first error is statistical and the second is systematic. }
    \label{tab:Results}
  \begin{center}
    \begin{tabular}{ccc}\hline\hline
    Sample                 & \yCP  & \deltaY \\
\KmKp      &  $( 1.60\pm0.46\pm 0.17)\%$ &  $( -0.40\pm0.44\pm 0.12)\%$\\
\pipi      &  $( 0.46\pm0.65\pm 0.25)\%$ &  $(\phantom{-}0.05\pm0.64\pm 0.32)\%$\\ \hline
Combined   &  $( 1.24\pm0.39\pm 0.13)\%$ &  $( -0.26\pm0.36\pm 0.08)\%$\\ 
  \hline\hline
    \end{tabular}
  \end{center}
\end{table}

\label{sec:crosschecks}

Various cross-checks have been performed to ensure that the
fit is unbiased and the assumptions in the fit model are well-founded. An offset in the resolution
function is measured in the fit to be $-4.75\pm0.51\fs$. This offset was
seen in our recent $\Km\pip$ mixing
analysis~\cite{Aubert:2007wf} and has also been observed in other
\babar{} measurements of charm decays. Because we measure
ratios of lifetimes, the presence of a common offset has minimal
impact on the values $\yCP$ and $\Delta Y$. However, differences in
the offset between the three decay modes, or between the \Dz\ and
$\Dzb$, could introduce a bias. No resolution offset is found in the 
MC samples. However, we are able to introduce offsets in the fits to
the MC sample of up to
twice the size of the offset in data by misaligning the Silicon Vertex
Tracker (SVT). In all
cases the offsets are found to be consistent between  all modes.

The fitting procedure has been validated with generic MC samples
weighted to the luminosity of the data sample and with dedicated
signal MC samples. The signal efficiency is found to be independent of
the true decay time and the fitted lifetimes are consistent with the
generated value. 

The assumption that the resolution function is the same for all decay
modes except for a scale factor is tested by fitting each sample
independently. This gives mixing parameters and resolution offsets
consistent with the nominal fit, but with significantly
larger statistical uncertainties. The lifetime has also been extracted
in independent fits to the flavor-separated samples of $\Dz\to\Km\pip$
and $\Dzb\to\Kp\pim$ decays. The fitted lifetimes and resolution
functions in these two samples are consistent with each other.

To cross-check the effect of the resolution offset, we performed  further studies by dividing the
data sample into subsamples with different sensitivities to
detector effects and fitting each subsample independently. Besides the
\Dst{} tagged samples used for this mixing measurement, we also use a
control sample of $\Dz\to\Kmpip$ decays where the \Dz is not required to come
from a \Dst{} decay. This untagged sample has about five times as many
\Dz decays as the \Dst{} tagged samples combined, allowing us to
divide the sample more finely. The quantities used to divide the data
into subsamples for these tests include the run period, the azimuthal and polar angle of the
\Dz meson, and the orientation of the \Dz decay plane with respect to
the X-Y (bending) plane of the detector. In all of the variables mentioned, the resolution
offset is observed to have a large variation (typically
between $-10\fs$ and $0\fs$), but the fitted lifetimes are 
consistent among samples. Furthermore, the weighted average of the
mixing parameters from the subdivided data samples is in almost all
cases nearly identical to that obtained by fitting the full data
sample with one common lifetime and resolution function as described
previously.  The largest
variation is observed with the polar angle of the \Dz meson in the laboratory
frame, where decays perpendicular to the beam line are found to have
almost no resolution offset, while decays into the forward region of
the detector have a large offset. Since the acceptance for $\Dz\to\KmKp$
decays is lower in the forward region than for $\Dz\to\Kmpip$ or
$\Dz\to\pipi$ decays, the polar angle dependence in the offset could
potentially introduce a different average offset for each of the three modes.
This is accounted for in the systematic errors.


\label{sec:systematics}

The systematic uncertainties on the mixing parameters are small since
most uncertainties in the lifetimes cancel in the ratios. 
We have considered variations in the 
signal and background fit models, changes to the event selection and
detector effects that could introduce biases in the
lifetime. Table~\ref{tab:SumYSystematicVariations} summarizes
the various systematic uncertainties. The evaluation of each of these is
described below. 
The systematic uncertainty on \yCP and \deltaY averaged over 
the two \CP modes is occasionally smaller than the individual uncertainties 
because of anti-correlations.

We vary the signal PDF shape, and the size and position of
the signal region.  As part of the PDF shape variations, we perform a fit
without a resolution offset. The effect of the polar angle dependence
in the resolution offset is evaluated by performing the fit with
separate, floating offsets in seven bins of polar angle, but sharing
all other resolution parameters and lifetimes across all polar angle
bins. The difference in the mixing parameters between this fit and the
nominal fit is found to be small ($<0.02\%$). The largest systematic
contribution to \yCPkk ($0.12\%$) is due to widening the signal region
mass interval from 15 to $25\mevcc$.  The choice of signal
region determines the level of mis-reconstructed signal events
included in the fit.

The mis-reconstructed charm background is a very small component in the
lifetime fit and is determined using MC events. Varying the
charm background fraction (depending on the mode) and the
effective lifetime, both within their associated uncertainties, yields a minor contribution to
the systematic uncertainty. 

Because of the high purity, the results
have little sensitivity to the modeling of the combinatorial
background, except in the \pipi mode where varying the fraction of combinatorial
background by 10\% yields a systematic uncertainty in \yCPpipi of $0.14\%$. We
also alter the fit procedure by using a different sideband region and by
substituting the MC decay time distribution for that obtained from fitting the
data. Neither variation contributes a large systematic uncertainty.

We have studied the effect of varying the event selection criteria,
which could potentially affect the lifetime measurement. 
Changing the treatment of events where multiple \Dstp\ candidates 
share one or more tracks (either keeping all of them or
throwing them all out) has little effect, while changing the upper
bound on the decay time uncertainty from 0.5 to 0.4\ps yields the largest systematic
uncertainty on \yCPpipi of $0.172\%$. As with the \Dz mass window, the
choice of the \terr{} range affects the level of mis-reconstructed
events.

To evaluate the effect of possible misalignments in the
SVT on the mixing parameters, 
signal MC events are reconstructed with different
alignment parameters, and the analysis is repeated.
The misalignments introduce 
resolution offsets in the MC of up to $10\fs$ and the corresponding fitted
lifetimes change by up to $3\fs$.  However, since the lifetimes of all
decay modes change
by similar amounts, the effect on \yCP and \deltaY is small. We also
changed the energy loss correction applied in the tracking by 20\%
since a previous analysis has shown that the energy loss is
underestimated in the reconstruction of data events~\cite{Aubert:2005gt}. This changes the
fitted lifetimes by about $0.5\fs$ but has little effect on the
mixing parameters.

\begin{table}[t]
  \caption{Summary of systematic uncertainties on \yCP and \deltaY, separately for $\KmKp$ and $\pipi$
and averaged over the two $CP$ modes, in percent.}
  \label{tab:SumYSystematicVariations}
  \begin{center}
  \begin{tabular}{lcccccccc}
    \hline
    \hline
&  &\multicolumn{3}{c}{$\sigma_{\yCP}$ (\%)}& &\multicolumn{3}{c}{$\sigma_{\deltaY}$ (\%)}\\

\cline{3-5}\cline{7-9}
Systematic & & $\KmKp$ & $\pipi$ & Av. & &$\KmKp$ & $\pipi$ & Av. \\\hline
Signal model          & &  0.130& 0.059     & 0.085     & &   0.072    &   0.265    &   0.062    \\
Charm bkg.        & &  0.062& 0.037     & 0.043     & &   0.001    &   0.002    &   0.001    \\
Combinatoric bkg.     & &  0.019& 0.142     & 0.045     & &   0.001    &   0.005    &   0.002    \\
Selection criteria           &  &  0.068& 0.178     & 0.046     & &   0.083    &   0.172    &   0.011    \\
Detector model        & &  0.064& 0.080     & 0.064     & &   0.054    &   0.040    &   0.054    \\\hline
 Quadrature sum       & &  0.172& 0.251     & 0.132     & &   0.122    &   0.318    &   0.083    \\\hline\hline
  \end{tabular}
  \end{center}
\end{table}

\label{sec:combination}

We combine the results shown in table~\ref{tab:Results}, with those from a previous
\babar\ study~\cite{Aubert:2003pz}, based on 91~\invfb of data, that does not require
a \Dstp{} parent to identify the \Dz decays.
While use of these untagged \Dz decays increases the sensitivity to \yCP
through a factor of five increase in statistics, it also introduces
different background behavior and therefore different systematic errors.
We have not used these untagged events in the current 
analysis, and thus the untagged data sample of the earlier analysis 
is essentially disjoint and its results statistically independent.
Systematic uncertainties in the previous analysis were dominated by
the limited number of simulated events.  Since the MC samples in the
present study are presented here are entirely independent, this
uncertainty is not correlated with those on the new results. Conservatively assuming
the remaining systematic uncertainties to be 100\% correlated, we combine 
the two results using the BLUE method~\cite{Lyons:1988rp} and obtain
$\yCP=[1.03\pm 0.33\stat \pm 0.19\syst)]\%$.


\label{sec:conclusions}

In summary, we have obtained a value of $\yCP=[1.24\pm 0.39\stat \pm
0.13\syst)]\%$ which is evidence of \Dz-\Dzb mixing at the $3\sigma$ level.
It is compatible with our previous result~\cite{Aubert:2003pz} and
the recent lifetime ratio measurement from Belle of
$\yCP=[1.31\pm 0.32\stat\pm 0.25\syst]\%$~\cite{Staric:2007dt}. 
We find no evidence for \CP violation and determine $\deltaY$ to be $[-0.26\pm
0.36\stat \pm 0.08\syst]\%$. The result is consistent with
SM estimates for mixing.

We are grateful for the 
extraordinary contributions of our \pep2\ colleagues in
achieving the excellent luminosity and machine conditions
that have made this work possible.
The success of this project also relies critically on the 
expertise and dedication of the computing organizations that 
support \babar.
The collaborating institutions wish to thank 
SLAC for its support and the kind hospitality extended to them. 
This work is supported by the
US Department of Energy
and National Science Foundation, the
Natural Sciences and Engineering Research Council (Canada),
the Commissariat \`a l'Energie Atomique and
Institut National de Physique Nucl\'eaire et de Physique des Particules
(France), the
Bundesministerium f\"ur Bildung und Forschung and
Deutsche Forschungsgemeinschaft
(Germany), the
Istituto Nazionale di Fisica Nucleare (Italy),
the Foundation for Fundamental Research on Matter (The Netherlands),
the Research Council of Norway, the
Ministry of Education and Science of the Russian Federation, 
Ministerio de Educaci\'on y Ciencia (Spain), and the
Science and Technology Facilities Council (United Kingdom).
Individuals have received support from 
the Marie-Curie IEF program (European Union) and
the A. P. Sloan Foundation.

\bibliographystyle{apsrev}

\begin{thebibliography}{21}
\bibitem[{foo()}]{footnote:CC}
\bibinfo{note}{The use of charge-conjugate modes is implied unless otherwise
  noted.}
\end{thebibliography}

\end{document}